\documentclass[showpacs, twocolumn]{revtex4-1}
\usepackage{graphicx}
\usepackage{amsmath}

\begin{document}

\title{Unconventional superfluids of fermionic polar molecules in a bilayer system}

\author{Abdel\^{a}ali Boudjem\^{a}a}

\affiliation{Department of Physics,  Faculty of Exact Sciences and Informatics, Hassiba Benbouali University of Chlef P.O. Box 151, 02000, Ouled Fares, Chlef, Algeria.}

\begin{abstract}
We study unconventional superfluids of fermionic polar molecules in a two-dimensional bilayer system with dipoles are head-to-tail across the layers.
We analyze the critical temperature of several unconventional pairings as a function of different system parameters. 
The peculiar competition between the $d$- and the $s$- wave pairings is discussed.
We show that the experimental observation of such unconventional superfluids requires ulralow temperatures,
which opens up new possibilities to realize several topological phases.

\end{abstract}


\maketitle


Recently, unconventional superfluids and superconductors where the gap parameters have symmetries different from the standard $s$-wave symmetry 
have attracted much attention because of their nontrivial statistical properties and topological behavior \cite{Volov, Mineev, Brus, Benn, Rice}. 

The advent of a quantum degenerate gas of polar molecule opens the door to a wide range of scientific explorations. 
Precision measurements, quantum-controlled chemical reactions and novel phases of matter \cite{Baranov, Pfau,Carr, Pupillo2012} are among a few prominent examples provided by 
an ultracold gas of polar molecules. 
Molecules have the capability to attain the transition dipole moment induced by a resonant microwave field, coupled with the internal rotational states.
Polar molecules may be better suited than neutral atoms for studying unconventional pairings, as they possess tunable electric dipole moment 
which can be induced by a static dc electric field, in addition to their own intrinsic dipole moment  \cite{KK, Aik, Deig, Kev}. 
An ideal platform to investigate the properties of polar molecules is a bilayer configuration, since it allows for stability against chemical reactions. 
Attractive interlayer interaction in such bilayer systems may also induce non-trivial interlayer pairings \cite{Pikov, Demler, Sarma, Baranov1, Zin, Bruun}.

\begin{figure}
\includegraphics[scale=0.8]{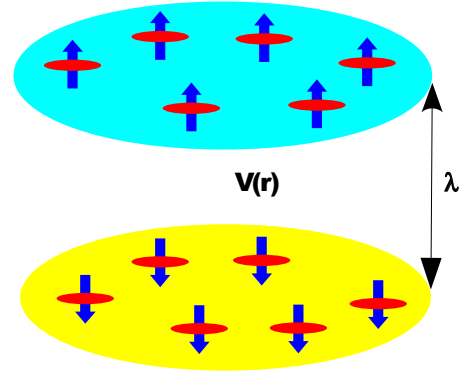}
\caption {(Color online) Bilayer system of cold polar molecules with dipoles oriented in opposite directions in different layers.}
  \label{sch}
\end{figure}

In this Letter we consider fermionic polar molecules loaded  in a two-dimensional (2D) bilayer geometry 
with dipole moments are aligned perpendicularly to the plane of motion and in opposite directions in different layers (see Fig.\ref {sch}). 
Such arrangement makes the interlayer interaction to be repulsive in short-ranged regime, while attractive in the long distance
leading to the emergence of novel interlayer superfluids, that are different to those obtained in \cite{Pikov, Demler, Sarma, Baranov1, Zin}. 
Most recently, the formation of a non-conventional $p$-wave superfluid in this configuration has been analyzed in \cite{Fedo}. 

Motivated by this, we study here various superfluid phases, that can possibly occur in the same bilayer system following the method described in  \cite{Fedo}.
We solve the $l$-wave interlayer scattering problem up to second order.
Then, we derive useful analytical expressions for the order parameter and the transition temperature for pairings 
through all angular momentum channels in terms of the interlayer spacing in the weak coupling regime. 
This procedure allows us to take into account a plethora of possibilities when calculating critical temperatures  of different competing 
orders as a function of the system parameters.
We will show in particular that the pairing symmetry is found to allow an "extended" $s$- ($s^*$) wave but markedly less so than the $d$-wave.
The interlayer $s^*$-wave pairing is indeed formally similar to the intersite $s^*$-wave in lattice \cite{Mic, Mic1}. 
The main difference between the two states is that in the case of lattice potential, the $s^*$-wave superfluid is determined by lattice symmetry, 
while in free case there is no special structure.
Note that an interesting early work by Micnas et \textit {al}. \cite{Mic, Mic1} dealing with high $T_c$ superconductor material through a finite range 
attractive interaction with repulsive on-site interaction relies on mean field theory, showed that close to half-filling, $d$-wave may dominate 
all other phases such as $s^*$-, $p$- and $g$-wave channels (see also e.g. \cite{Buk, Jas}).
Superfluids with a $d$-wave order parameter are particularly interesting and exhibit a wealth of fascinating properties that are highly sought after for nanoscience applications.
They have peculiar connections with high $T_c$ superconductivity \cite{Laugh} and spin quantum Hall fluid phase \cite{Fisher}. 
The experimental realization and detection of such unconventional pairings  are discussed.

Let us consider the bilayer system of Fig.\ref {sch}, where the two layers are separated by a distance $\lambda$.
Polar molecules have permanent electric dipole moment $d$ which establish a long-range and anisotropic interaction among molecules. 
The Hamiltonian of the system reads
\begin{align}\label{ham}
\hat H &=\sum_a\int d {\bf r } \, \hat \psi_a^\dagger ({\bf r}) \left(-\frac{\hbar^2}{2m}\Delta-\mu\right)\hat\psi_a(\mathbf{r}) \\
&+\frac{1}{2}\sum_{a,b}\int d {\bf r } d {\bf r'}\, \hat\psi_a^\dagger({\bf r}) \hat\psi_b^\dagger ({\bf r'}) V({\bf r-r'})\hat\psi_b ({\bf r'}) \hat\psi_a ({\bf r}),
\nonumber
\end{align}
where $a=1,2$ is the layer index, $\hat\psi_a^\dagger$ and $\hat\psi_a$ denote, respectively the usual creation and annihilation field operators and $\mu$ is the chemical potential.
The dipole-dipole interparticle interaction potential $V(r)$ may split into two parts: the interlayer interaction potential ($a \neq b$) is
\begin{equation} \label{pot}
V(r) = - d^2\frac{r^2-2\lambda^2}{(r^2+\lambda^2)^{5/2}}.
\end{equation}
At large distance $r$, the potential (\ref{pot}) is attractive which may lead to the formation of an interlayer bound state.
The Fourier transform of the interlayer potential reads
\begin{align} \label{FT}
& V({\bf q}) = \int d{\bf r} V ({\bf r}) e^{- i {\bf q r}} = \frac{2\pi\hbar^2} {m} r_*|{\bf q}| e^{-|{\bf q}|\lambda}, 
\end{align}
where $r_*=md^2/\hbar^2$ is the characteristic dipole-dipole distance and $m$ is the particle mass. \\
The Fourier transform of the intralayer interaction ($a = b$) can be readily achieved by putting $\lambda=0$ in Eq. (\ref{FT}) which gives $V({\bf q})=(2\pi\hbar^2/m) r_*|{\bf q}|$.



\begin{figure}
\includegraphics[scale=0.8]{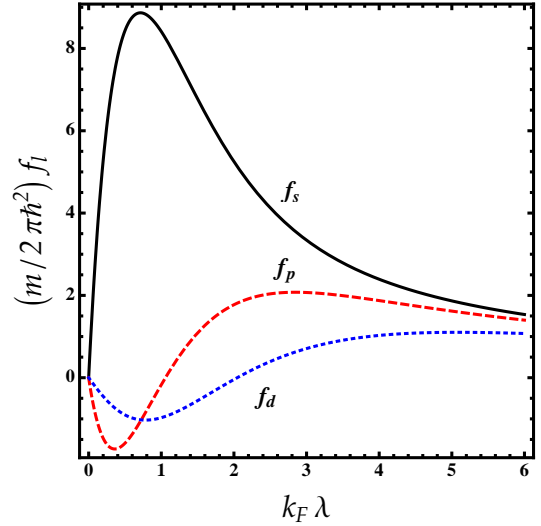}
\caption {(Color online) Born approximation for the $s$-, $p$- and $d$-wave scattering amplitude from Eq.(\ref{Born}) as a function of $k _F\lambda$ for $r_*/\lambda=3$.}
  \label{amplf}
\end{figure}

In the first Born approximation, the scattering amplitude is given by
\begin{align}\label{Born}
f_l ^B ({\bf k,k'}) =  \int d{\bf r} V ({\bf r}) e^{i {\bf (k-k') r}}.
\end{align} 
In order to obtain the $l$-wave part of the scattering amplitude, we multiply Eq.(\ref{Born}) by $e^{-i \phi l}$ and integrate over $d\phi/2\pi$, 
where $\phi$ is the angle between the vectors $k$ and $k'$.
In Fig.\ref{amplf} we plot on-shell amplitudes of the $s$-, $p$- and $d$-wave scattering at $k = k_F$, where $k_F$ is the Fermi momentum.
As is seen the first Born approximation reveals that the $s$-wave scattering amplitude $f_s$ is always positive 
preventing the occurrence of the standard $s$-wave interlayer pairing.
Whereas $f_p$ and $f_d$ associated respectively with $p$-wave and $d$-wave are negative. 
Importantly, for $k \lambda \leq 0.7$, the $d$-wave scattering amplitude becomes dominant ($f_d > f_p$). 
Therefore, an interlayer $d$-wave superfluid is possible at very low temperature.

An accurate analysis of the gap equation requires the inclusion of higher-order (up to the second order) contributions to the scattering amplitude. \\
The off-shell $l$-wave scattering amplitude is given by 
\begin{equation}\label{scat}
f_l (k',k)=\int_0^{\infty}  J_l (k'r) V(r) \psi_l (k,r)  2\pi r dr,  
\end{equation}
where $J_l$ is the Bessel function and $\psi_l (k,r)$ is the true wavefunction of the $l$-wave relative motion with momentum $k$. 
It is normalized in such a way that for $r \rightarrow \infty $, one has $\psi_l (k,r)=J_l (kr)- i (m/4\hbar^2) f(k)  H_l (kr)$, with $H_l$ being the Hankel function.
Using this relation, the  off-shell scattering amplitude (\ref{scat}) takes the form \cite{Zhen}
\begin{equation}\label{scat1}
f_l (k',k)=\frac{\bar f_l(k',k)}{1+i  (m/4\hbar^2) \bar f_l(k)},
\end{equation}
where $\bar f_l=\bar f_l^{(1)}+\bar f_l^{(2)}$ is real and follows from (\ref{scat}) with
\begin{align}\label{scat2}
\bar f_l^{(1)} (k'\lambda,k\lambda)& =\frac{2\pi\hbar^2} {m}  \frac{k r_*}{k \lambda} \int_0^{\infty} x d x J_l(k' \lambda x) J_l(k \lambda x) \nonumber \\
& \times \frac{x^2-2}{(x^2+1)^{5/2}},
\end{align} 
is the first order correction to off-shell scattering amplitude. 
For the second order contribution, we need only the on-shell form for the solution of the gap equation 
\begin{align}\label{scat3}
\bar f_l^{(2)}(k\lambda)& =\frac{2\pi^2\hbar^2} {m} \frac{(k r_*)^2}{(k \lambda)^2} \int_0^{\infty} x d x J_l^2(k \lambda x) \frac{x^2-2}{(x^2+1)^{5/2}} \nonumber \\
& \times \int_x^{\infty} y d y J_l (k \lambda y) N_l (k \lambda y) \frac{y^2-2}{(y^2+1)^{5/2}},
\end{align}
where $N_l$ is the Neumann function.


For a weak interlayer attractive interaction, the gap equation for the momentum-space order parameter reads
\begin{align} \label{gap}
\Delta ({\bf k})&= - \int \frac{d^2k'}{(2\pi)^2} V ({\bf k'- k})\Delta ({\bf k'}) \frac{ \text{tanh}\left(\varepsilon_{k'}/2T \right)} {2\varepsilon_{k'}},
\end{align}
where $\varepsilon_k=\sqrt{(E_k-\mu)^2+|\Delta (k)|^2}$ and $E_k=\hbar^2k^2/2m$ is the energy of free particle.\\
Expressing the Fourier transform of the interaction potential, $V ({\bf k'- k})$, in terms of the off-shell scattering amplitude $f_l ({\bf k, k'})$ \cite{Landau}.
Then using the expansions $\Delta ({\bf k})=\sum_l \Delta (k) e^{i \phi_k l}$ and  $f_l ({\bf k},{\bf k'})=e^{i (\phi_k-\phi_{k'}) l} f_l (k,k')$ \cite{Zhen}.
After having multiplying both sides of the resulting equation by $e^{-i \phi_k l}$ and integrating over $d\phi_{k}$ and $d\phi_{k'}$, the renormalized gap equation takes the form

\begin{align} \label{gap1}
\Delta (k)&= - {\cal P}\int \frac{d^2k'}{(2\pi)^2} f_l(k', k)\Delta (k') \\
& \times  \left[\frac{ \text{tanh}\left(\varepsilon_{k'}/2T \right)} {2\varepsilon_{k'}}-\frac{1}{2(E_{k'}-E_k-i0)}\right], \nonumber 
\end{align}
where ${\cal P}$ is the principal part of the integral.\\
Equation (\ref{gap1}) may be used for calculating $\Delta (k)$ and the superfluid transition temperature $T_c$. 
In 2D Fermi gas, the transition from the normal to superfluid phase is of the Kosterlitz-Thouless type. However, in weakly interacting regime,
the Kosterlitz-Thouless transition temperature is very close to $T_c$ and it can be obtained within the BCS approach \cite {Miy}.

\begin{figure}  [htbp]
\centerline{
\includegraphics[scale=0.8,clip]{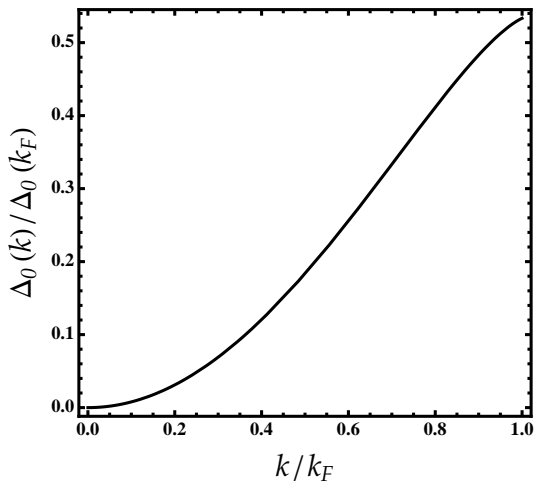}}
 \caption{The zero-temperature gap $\Delta_0 (k)$ as a function of $k/k_F$.}
\label{GT0} 
\end{figure}

The main contribution to the integral on the right-hand side of Eq.(\ref{gap1}) is related to the first term in square brackets and
comes from a narrow vicinity of the Fermi surface. Singling out this main contribution, we find a direct relation between $\Delta (k)$ and $\Delta (k_F)$ 
\begin{equation} \label{gap2}
\Delta (k)= \Delta (k_F)  \frac{\bar f_l (k_F,k)} {\bar f_l (k_F)}.
\end{equation}
Inserting Eq.(\ref{gap2}) into (\ref{gap1}), makes it possible to obtain a relation between the zero-temperature order parameter on the Fermi surface $\Delta_0 (k_F)$
and $T_c$.
The behavior of the zero-temperature gap $\Delta_0 (k)$ in units of $\Delta_0 (k_F)$ is displayed in Fig.\ref{GT0}.

To calculate the superfluid critical temperature, we solve the gap equation (\ref{gap1}) at $k=k_F$.   
So, we divide the region of integration into two parts:$ |E_k-E_F| < \omega $ and $ |E_k-E_F| > \omega$. 
In the first region the main contribution to $\Delta (k)$ comes from $k'$ close to $k_F$.
In the second region, $ |E_k-E_F| > \omega$, we put $\text{tanh}\left(\varepsilon_{k'}/2T \right)/ (2\varepsilon_{k'}) = 1/(2|E_{k'}-\mu|)$
and  retaining only the leading low-momentum contribution to the off-shell amplitude $\bar f_l (k'_F,k)$ (for more details see Ref.\cite{Levin}). 
Then, summing all the contributions by canceling $\Delta (k_F)$ and setting $E_F \sim \omega$ in the terms proportional to $(k_F r_*)^2$, we obtain 
\begin{align} \label{gap3}
1= &(k_F r_*) F_l^{(1)} (k_F\lambda ) \ln\left( \frac{2e^{\gamma -h_l (k_F\lambda )}} {\pi} \frac{E_F}{T_c} \right)  \nonumber \\
& - (k_F r_*)^2 F_l^{(2)} (k_F\lambda ) \ln\left( \frac{E_F}{T_c} \right), 
\end{align}
where $\gamma=0.5772$ is the Euler constant, $F_l^{(1)}=\rho \bar f_l^{(1)}/(kr_*)$, $F_l^{(2)}=\rho \bar f_l^{(2)}/[\pi(kr_*)^2]$, $\rho=m/(2\pi \hbar^2)$ is the density of state
on the Fermi surface, and the function $h_l$ is defined as
$$h_l(k_F\lambda )=-2\int_0^1 \frac{xdx}{1-x^2} \left\{\left[ \frac {F_l^{(1)} (k_F\lambda, k_F\lambda x)} {F_l^{(1)} (k_F\lambda)} \right]^2-1\right\}.$$
The effective mass has to be replaced instead of the bare mass in the gap equation  using the Fermi liquid theory \cite{Landau}
$$ \frac{1}{m}=\frac{1}{m^*} +\frac{1}{(2\pi \hbar)^2} \int_0^{2\pi} F_l(\theta) \cos(\theta) d\theta,$$
where $F_l(\theta)= 2\bar f_l (k_F |\sin (\theta/2)|)$ on the Fermi surface. \\
Working in the limit $k_F r_* \ll1$, the critical temperature turns out to be given:
\begin{equation} \label{Tcrt}
T_c= E_F A_l(k_F\lambda )  \exp \left[-\frac{1}{k_F r_*F_l^{(1)}(k_F\lambda ) }\right],
\end{equation} 
where 
\begin{equation}   \label{gap}
 \begin{split}
A_l(k_F\lambda ) = \exp \left [\gamma+\ln (2/\pi)-h_l(k_F\lambda ) - \frac{F_l^{(2)} (k_F\lambda )}{F_l^{(1)2}(k_F\lambda ) } \right.  \\
\left. - \frac{4}{ \pi  (4 l^2-1) F_l ^{(1)}(k_F\lambda)} \right].  \nonumber
\end{split}
\end{equation}
Expression (\ref{Tcrt}) enables us to determine straightforwardly the transition temperature for any $l$-wave superfluid as a funcion of the interlayer spacing.
For $l=1$, Eq.(\ref{Tcrt}) excellently agrees with that obtained in \cite{Fedo} for the $p$-wave superfluid.

\begin{figure}  [htbp]
\centerline{
\includegraphics[scale=0.8,clip]{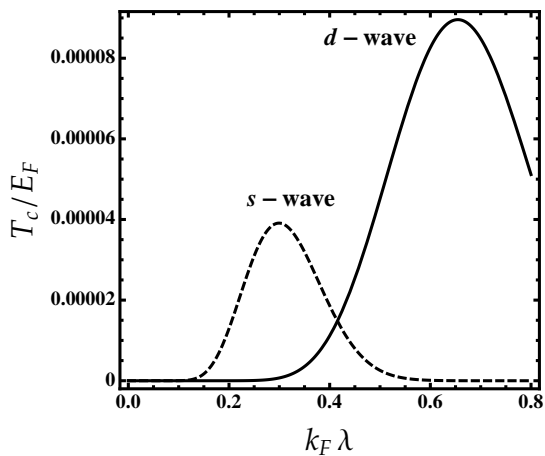}}
 \caption{Critical temperature $T_c$ as a function of $k_F\lambda$.}
\label{Tc} 
\end{figure}

In Fig.\ref{Tc} we show the critical temperature for $s$- and $d$-wave superfluids.
The $s$-wave pairing is obtained numerically by solving the full BCS equation (\ref{gap1}) with momentum dependence in the pairing wavefunction.
Many intersting  results presented in Fig.\ref{Tc} should be noted. 
First, the $T_c$ of both pairings decreases rapidly for $k_F\lambda >$ 0.4 and 0.8, respectively due to the fast decay of their corresponding  scattering amplitudes.
The optimal values of $k_F\lambda$ are around $0.3$ and $0.65$  with $T_c$ reaching values on the order of $\sim 4 \times 10^{-5} E_F$ and $\sim 9 \times 10^{-5} E_F$
for $s$- and $d$-wave pairings, respectively. 
Second, a closer look tells us that the $d$-wave become dominant for larger $k_F\lambda$, 
while the $s$-wave state is favored only in lower density (smaller $k_F\lambda$) regime.
Both pairings coincide at $k_F\lambda=0.42$ and thus, have the same  transition temperature $T_c=2 \times 10^{-5} E_F$.
Because these pairings are supported by Fermi surface scattering enhancements, the $s$-wave interaction gains a positive value except for very small momenta.
This scenario should be contrasted with the $d$-wave pairing, in which the $f_d$ is negative and large in amplitude. 
These are the key reasons why the $d$-wave pairing is favored over the $s^*$-wave pairing channel,
although both pairing symmetries are possible contenders for unconventional superfluids of fermionic polar molecules.
Note that the critical temperature of both states can be enhanced if one takes into account the Gor'kov and Melik-Barkhudarov corrections.

We now discuss the  experimental issues related to the realization and detection of unconventional polar superfluids proposed in this paper.
For ${}^6$Li${}^{133}$Cs molecules ($r_*=6 \times 10^5 a_0$ \cite{Jin} with $a_0$ being the Bohr radii) placed in a bilayer system (created by a 1D subwavelength lattice) with
an interlayer separation $\lambda =50$ nm \cite{Fedo, Gonz,Yi}, one achieves densities $n \sim 8\times 10^8$ cm$^{-2}$ in each layer for $k_F \lambda=0.65$.
In this case, the $d$-wave superfluid transition temperature is around $T_c \sim 665$ pk.
Whereas, for the $s$-wave pairing,  $T_c$ reaches $\sim 63 $ pk for $k_F \lambda=0.3$ which is arround ten times smaller than that of the $d$-wave channel.
Note that, Leanhardt et \textit {al}. \cite{Kitt} had already succeeded to cool a BEC of sodium atoms at temperature in the order of 450$\pm$ 80 pk, 
using adiabatically a decompression process and subsequent evaporative cooling of partially condensed atomic vapours in a very shallow gravito-magnetic trap.


The critical temperature strongly depends on the interlayer spacing. 
For $\lambda \rightarrow 0$ i.e. one layer system, the $d$-wave ($l=2$) critical temperature simplifies to $T_c= E_F A_2  \exp \left[-15/(4 k_F r_*)\right]$,
where  $A_2 = \exp [\gamma+\ln (2/\pi)-h_2 -15\pi^2/128-5]$ and $h_2 \sim 0.88$.
It is clear that for the same density, the critical temperature in one layer is larger than in the bilayer system due to the small exponent parameter.


The production of a quantum degenerate gas of polar molecules trapped in a 2D optical lattice may provide a versatile land of reaching these ultralow temperatures.
One possibility to create such an ensemble is to directly image individual molecules and use that information to make a much lower entropy gas of molecules, 
and hence lowered the corresponding temperature \cite{PC}. 
Note that the creation of  low-entropy quantum degenerate gas of polar molecules in a 3D optical lattice has recently been reported in \cite{Stev, Jacob},
where effects on pairing that arise from $d$-wave Feshbach resonance is observed.



In conclusion, we have studied various exotic superfluids of fermionic polar molecules residing in 2D bilayer geometry 
where dipole moments are aligned head-to-tail across the layers.
Within the BCS approach, we have calculated the critical temperature of different unconventional pairings as a function of system parameters.
Our results indicate that these polar molecule ensembles occur at very colder temperatures.
The competition among the $d$-wave and the $s^*$-wave superfluids has been analyzed.
In the strong coupling regime, the dimerization transition is still possible and should be investigated by means of sophisticated Monte carlo simulations.
We hope that the insights obtained in this work offer intriguing perspectives to the exploration of new physics phenomena that occur only in very low energy scales.  
Understanding such anisotropic superfluids would also open new prospects for precision measurements of fundamental constants (due to the much slower atoms) \cite{Zel}.


We are grateful to Gora Shlyapnikov for contributions at the early stage of the work.
We thank Sergey Matveenko, Mikhail Baranov, Wolfgang Ketterle, Luis Santos, Matthias Weidem\"uller and Jun Ye for valuable discussions.
We acknowledge support from the University of Chlef.
We would like to thank the LPTMS-France for a visit, during which this work was conceived.

\end{document}